\documentclass{llncs}
\usepackage{graphicx}
\usepackage{geometry}
\usepackage{array}
\usepackage{lscape}
\usepackage{tabularx}
\usepackage{booktabs} 
\usepackage{makeidx}  % allows for indexgeneration
\begin{document}

\mainmatter              % start of the contributions
\title{ Using 3D Scan to Determine Human Body Segment Mass in OpenSim Model}

\author{Jing Chang\inst{1}\and
Damien Chablat\inst{2}\and
Fouad Bennis\inst{1} \and
Liang Ma\inst{3}} 
\authorrunning{Jing Chang et al.} % abbreviated author list (for running head)

\institute{Ecole Centrale de Nantes, Laboratoire des Sciences du Num\'erique de Nantes (LS2N), UMR CNRS 6004, 44321 Nantes, France
\and
CNRS, Laboratoire des Sciences du Num\'erique de Nantes \\
(LS2N), UMR CNRS 6004, 44321 Nantes, France\\
\and
Department of Industrial Engineering, Tsinghua University,
Beijing, 100084, P.R.China\\
\email{\{Jing.chang, Damien.Chablat, Fouad.Bennis\} @ls2n.fr\\
liangma@tsinghua.edu.cn}}
\maketitle              % typeset the title of the contribution

\begin{abstract}
Biomechanical motion simulation and dynamic analysis of human joint moments will provide insights into Musculoskeletal Disorders. As one of the mainstream simulation tools, OpenSim uses proportional scaling to specify model segment masses to the simulated subject, which may bring about errors. This study aims at estimating the errors caused by the specifying method used in OpenSim as well as the influence of these errors on dynamic analysis. A 3D scan is used to construct subject's 3D geometric model, according to which segment masses are determined. The determined segment masses data is taken as the yardstick to assess the errors of OpenSim scaled model. Then influence of these errors on the dynamic calculation is evaluated in the simulation of a motion in which the subject walks in an ordinary gait. Result shows that the mass error in one segment can be as large as 5.31\% of overall body weight. The mean influence on calculated joint moment varies from 0.68\% to 12.68\% in 18 joints. 

In conclusion, a careful specification of segment masses will increase the accuracy of the dynamic simulation. As far as estimating human segment masses, the use of segment volume and density data can be an economical choice apart from referring to population mass distribution data.

\end{abstract}
\section{Introduction}

Musculoskeletal Disorders (MSDs) makes up the vast proposition of the occupational diseases \cite{Eurogip}.  Inappropriate physical load is viewed as a risk factor of MSD \cite{Chaffin1999}. Biomechanical analysis of joint moments and muscle loads will provide insight into MSDs. 

Over the past decades, many tools have been developed for biomechanical simulation and analysis. OpenSim \cite{Delp2007} is a virtual human modeling software that has been widely used for motion simulation and body/muscle dynamic analysis \cite{Thelen2006,Kim2017}. The simulation in OpenSim should be based on a generic virtual human that consists of bodies, muscles, joint constraints, etc., as shown in Figure 1.

\begin{figure}[htbp]
  \centering
  \includegraphics[width=0.4\textwidth]{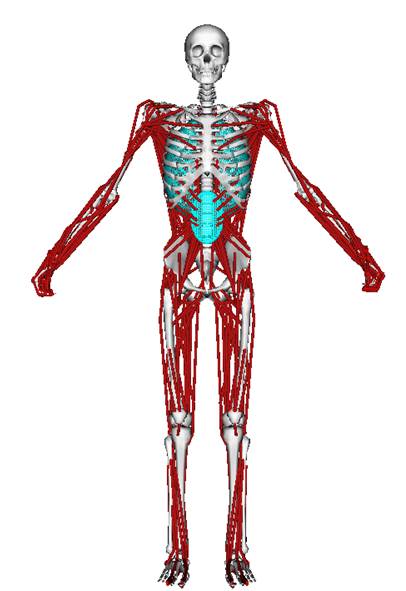}\\
  \caption{A generic OpenSim model.}
  \label{Fig1}
 \end{figure}

A simulation is generally started by scaling the generic model specifically to the subject’s geometric and mass data. The subject’s body geometric data is obtained using a motion capture system, which records the spatial positions of flash reflecting markers that attached to the specific locations of the subject; then the generic OpenSim model is adjusted geometrically with attempts to minimize the position deviations between virtual markers and corresponding real markers. This makes a subject-specific model out of the generic model.

The geometrical adjustment increases the accuracy of posture simulation and kinematic analysis that follow. For accurate dynamic analysis, the body segment inertial parameters, such as segment masses, should also be adjusted specifically to each subject. In OpenSim, this adjustment is carried out by scaling the mass of each segment of the generic model proportionately with respect to the whole body mass of the subject. 

This method of determining segment mass is based on the assumption that the mass distribution among body segments is similar among humans, which is not always the case. For example, the mean mass proportion of the thigh has been reported to be 10.27\% \cite{Clauser1969}, 14.47\% \cite{DeLeva1996}, 9.2\% \cite{okada1996}, and 12.72\% \cite{Durkin2003}, which indicates significant individual difference. Therefore, the scaling method used by OpenSim is likely to cause errors in the following dynamic analysis. There is a necessity to estimate the errors.

This paper aims at estimating the errors caused by the scaling method used in OpenSim. Firstly, subject's segment masses are determined based on the accordingly 3D geometric model constructed with the help of 3D scan. The determined data is taken as an approximation to the true value of the subject's segment mass.  Secondly, this set of data, as well as the proportionately scaled segment mass data, is used to specify a generic OpenSim model. Errors caused by proportionately scaling are calculated. Finally,  influence of the errors on dynamics analysis is checked on a simulation of a walking task. 

The method to approximate subject's segment masses, model specification and dynamic simulation are described in chapter 2. Results are presented in chapter 3. These results are then discussed in chapter 4, followed by a conclusion in Chapter 5.
\section{Methodology}
\subsection{Approximating segment masses with 3D scan}
%
%%% 3D scan %%%
A whole-body 3D scan was conducted to a male subject (31 years old, 77.0~kg, 1.77~m) with a low-cost 3D scanner (Sense$^{TM}$ 3D scanner). Before scanning, reflecting markers were placed on the subject to notify the location of each joint plate as shown in Figure 2. The locations of the joint plates were set according to Drillis (1966) \cite{Drillis1966}, which meant to facility the dismemberment of the 3D model. During scanning, careful caution was taken to make sure that no extra contact between limbs and the torso. The scanned 3D model was stored in a stl mesh file, as shown in Figure~3.

\begin{figure}[htbp]
  \centering
  \includegraphics[width=0.3\textwidth]{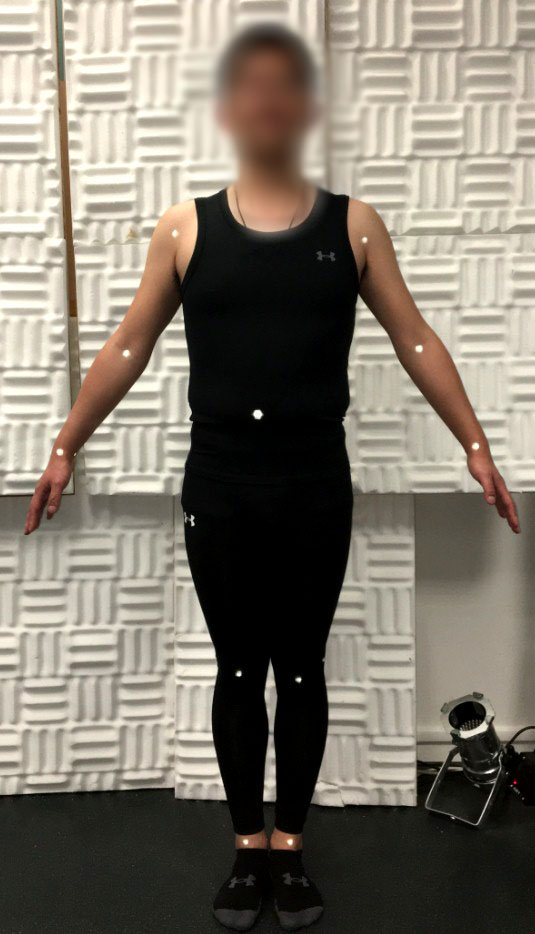}\\
  \caption{Body markers that indicate the location of joint plates.}
  \label{Figure 3}
 \end{figure}
 \begin{figure}[htbp]
  \centering
  \includegraphics[width=0.6\textwidth]{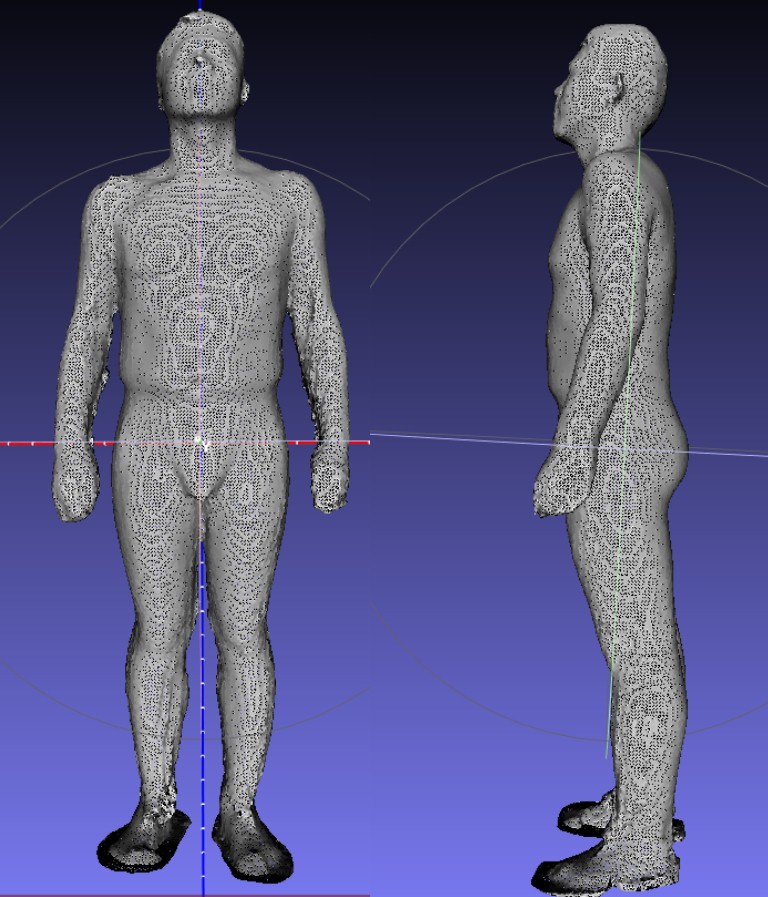}\\
  \caption{The 3D geometric model of the subject generated by 3D scan.}
  \label{Mesh}
 \end{figure}
 
Then the 3D model was dismembered into 15 parts in the way that described by Drillis (1969) \cite{Drillis1966}. Body markers and body parts lengths are referred. An example of the dismembered body part (Pelvis) is shown in Figure~4. Then, the volume of each body part was calculated.
 
\begin{figure}[htbp]
  \centering
  \includegraphics[width=0.3\textwidth]{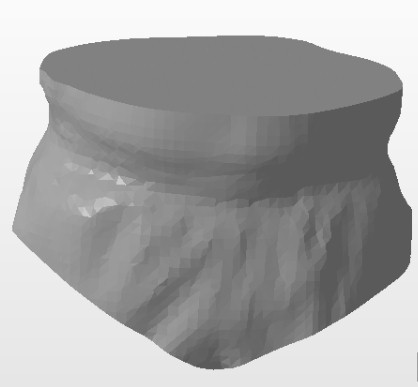}\\
  \caption{The mesh of pelvis dismembered from the whole-body 3D model.}
  \label{Figure 4}
 \end{figure}
   
To analyze the results obtained,  the water displacements of eight distal body parts (hands, lower arms, feet, lower legs) were also measured, as described by Drilis (1966) \cite{Drillis1966}. 
%%%%%%%%%%%%%%%%%%%%%%%%%%%%%%%%%%%%%%%%%%%%%%%%%%%%%%%%%%
\subsection{Specification of OpenSim models to the subject}
In this step, a generic OpenSim model is specified  to our subject in aspect of body segment mass. The model is developed by Delp S.L. et al (1990) \cite{Delp1990} (http://simtk-confluence.stanford.edu:8080/display/OpenSim/Gait+2392+and+\\2354+Models). It consists of 12 bodies, 23 degrees of freedom, and 52 muscles. The unscaled version of the model represents a subject that is about 1.8~m tall and has a mass of 75.16~kg. 
\begin{figure}[htbp]
  \centering
  \includegraphics[width=0.6\textwidth]{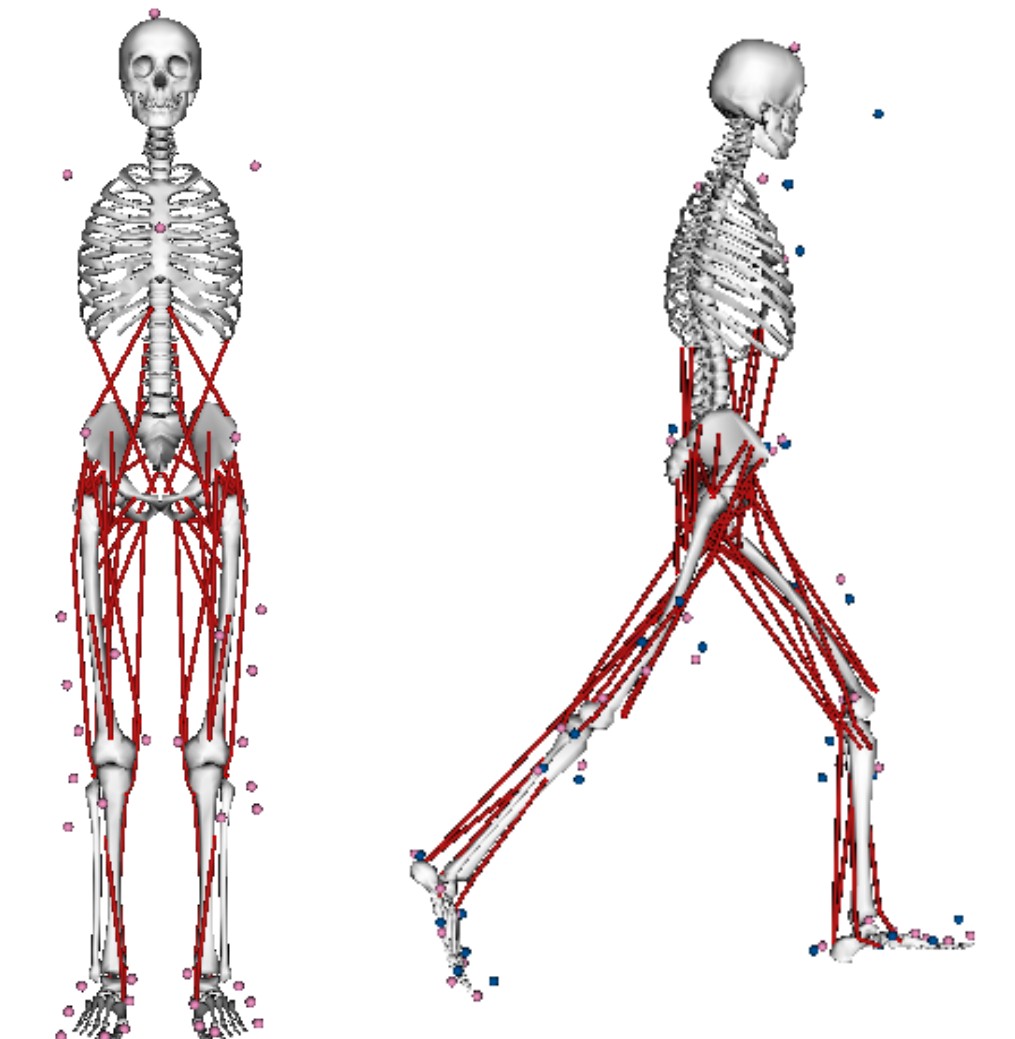}\\
  \caption{The OpenSim model used for error analysis.}
  \label{Figure 5}
 \end{figure}
The approximating body segment mass data obtained from the process in chapter 2.1 as well as the proportionately weight-scaled body segment mass data is used to specific the generic model. The former is considered as the yardstick to estimate the error of the latter. 
\subsection{Dynamic simulation in OpenSim}
An simulation is conducted on the two specific models. Simulation data comes from previous researches \cite{John2012}. The subject walks two steps in 1.2~s in an ordinary gait. Data of Spatial posture is collected at a frequency of 60~Hz and the and ground reaction forces at a frequency of 600~Hz. Inverse dynamic analysis is conducted on both models. Joint moments are calculated and compared between the two models.  

\section{Results}
\subsection{The estimation of body segment masses}%

Significant difference was found between volumes calculated from the 3D scanned geometric model and that measured by water displacement. For lower leg, the difference is as large as 27\% (4.5~l vs. 3.3~l). To approximate the real segment masses, assumption is made that the volume distribution of the 3D model merged by 3D scan among head, torso, pelvis, and upper limbs is the same as that of the real subject. Density data of the body parts \cite{Wei1995} were used to calculate the whole-body density of the subject, which, as well as body weight, gives estimation of whole-body volume. Then the overall volume was distributed to each segment with respect to the relative volume ratio of the 3D geometric model. In this way, segment volumes and masses are approximated. 

Relative data is shown in Figure~6. The whole-body volume calculated from the 3D geometric model is 7.31\% larger than the estimated whole-body volume (81.81~l to 76.24~l). The whole body density is estimated to be 1.01~g/ml. Mass proportion of the thigh is about 11.30\%, which is between that reported by  Clauser et al., (1969) (10.27\%) \cite{Clauser1969}, Okada (1996) (9.2\%) \cite{okada1996} and by De Leva (1996) (14.47\%) \cite{DeLeva1996},  Durkin \& Dowling (2003) (12.72\%) \cite{Durkin2003}.

\newgeometry{left=4cm,bottom=2.5cm}

\begin{landscape}

\begin{table}
\centering
  \caption{Volume, density and mass of the whole body and segments.}
  {\begin{tabular}{p{3.5cm}|p{1.3cm}p{1.2cm}p{1.2cm}p{1.2cm}p{1.2cm}p{1.2cm}p{1cm}p{1cm}p{1cm}p{1cm}p{1cm}p{1cm}p{1.2cm}p{1.2cm}p{1.2cm}p{1.2cm}}
  \toprule
  {} & {\noindent{\textbf{Overall}}} & {\noindent{\textbf{Pelvis}}} & {\noindent{\textbf{Head}}}& {\noindent{\textbf{Torso}}} & {\noindent{\textbf{Upper arm-l}}} & {\noindent{\textbf{Upper arm-r}}} & {\noindent{\textbf{Lower arm-l}}} & {\noindent{\textbf{Lower arm-r}}} & {\noindent{\textbf{Upper leg-l}}}& {\noindent{\textbf{Upper leg-r}}}& {\noindent{\textbf{Lower leg-l}}}& {\noindent{\textbf{Lower leg-r}}}& {\noindent{\textbf{Foot-l}}}& {\noindent{\textbf{Foot-r}}}& {\noindent{\textbf{Hand-l}}}& {\noindent{\textbf{Hand-r}}}\\
\midrule
Volume(l)-3D scanned model &81.81&11.29&5.65&29.90&1.63&1.72&1.38&1.34&8.74&8.65&4.42&4.53&1.06&0.89&0.49&0.58\\
Volume(l)-water displacement& & & & & & &1.01&1.12& & & 3.25&3.35&1.00&1.00&0.46&0.46\\
Volume(l)-estimated& 76.24 & 10.81 & 5.41&28.64&1.56&1.65&1.01&1.12&8.37&8.29&3.25&3.35&1.00&1.00&0.46&0.46 \\
Density(kg/l)\cite{Wei1995}&1.01&1.01&1.07&0.92&1.06&1.06&1.10&1.10&1.04&1.04&1.08&1.08&1.08&1.08&1.11&1.11\\
Estimated mass(kg) & &10.92&5.79&26.35&1.66&1.75&1.11&1.23&8.71&8.62&3.51&3.62&1.08&1.08&0.51&0.51\\
\bottomrule
\end{tabular}}
 \label{tab1}
\end{table}

\begin{table}[htbp]
\centering
 \caption{Errors of proportionally scaled segment masses with respect to the approximate masses.}
  {\begin{tabular}{m{4cm}|m{1.5cm}p{1.5cm}p{2cm}p{2cm}p{2cm}p{2cm}p{1.5cm}p{2cm}}
  \toprule
  {} & {\noindent{\textbf{Pelvis}}} & {\noindent{\textbf{Torso}}} & {\noindent{\textbf{Upper leg-l}}}& {\noindent{\textbf{Upper leg-r}}}& {\noindent{\textbf{Lower leg-l}}}& {\noindent{\textbf{Lower leg-r}}}& {\noindent{\textbf{Talus}}}& {\noindent{\textbf{Calcaneus}}}\\
\midrule
Approximate mass(kg)&10.92&38.91&8.71&8.62&3.51&3.62&0.07&0.87\\
Proportionally scaled mass (kg)&11.98 &34.82 &9.46 &9.46 &3.76 &3.76 &0.10 &1.28\\
Absolute error (kg)&1.06 &-4.08 &0.75&0.84&0.25&0.15&0.03&0.41 \\
Percentage of absolute error in overall body weight&1.37\% &-5.30\%&0.98\%&1.10\%&0.33\%&0.19\%&0.04\%&0.53\%\\
Relate error&9.66\%&-10.49\%&8.67\%&9.80\%&7.26\%&4.06\%&47.42\%&47.42\%\\

\bottomrule
\end{tabular}}
 \label{tab2}
\end{table}

\end{landscape}
\restoregeometry

\subsection{Errors analysis of the OpenSim scaled model specific to the subject}%

Segment masses generated by proportionately scale and 3D modeling are used to specify the OpenSim generic model, which bring about two specific models (noted as scaled model and approximate model). Figure~7 shows the segment mass data of the two models.  Errors of the scaled model segment mass are between 4.06\% and 47.42\%. The most significant error merges from foot data which, however, represents only a small part of the overall body mass.  

\subsection{Motion simulation and dynamic analysis}%
Both the proportionately scaled segment mass data and approximate segment mass data are used to specify the OpenSim generic model, bringing about two specific models (noted as scaled model and approximate model).
   
Motion simulation is conducted on the two models. The simulated motion includes two steps of walking, lasting for 1.2~s. Since the two models differ in only segment mass, no difference in kinematic analysis is shown. As an example, the angles, velocity and acceleration of right hip flexion is shown in Figure~8.

\begin{figure}[htbp]
  \centering
  \includegraphics[width=0.8\textwidth]{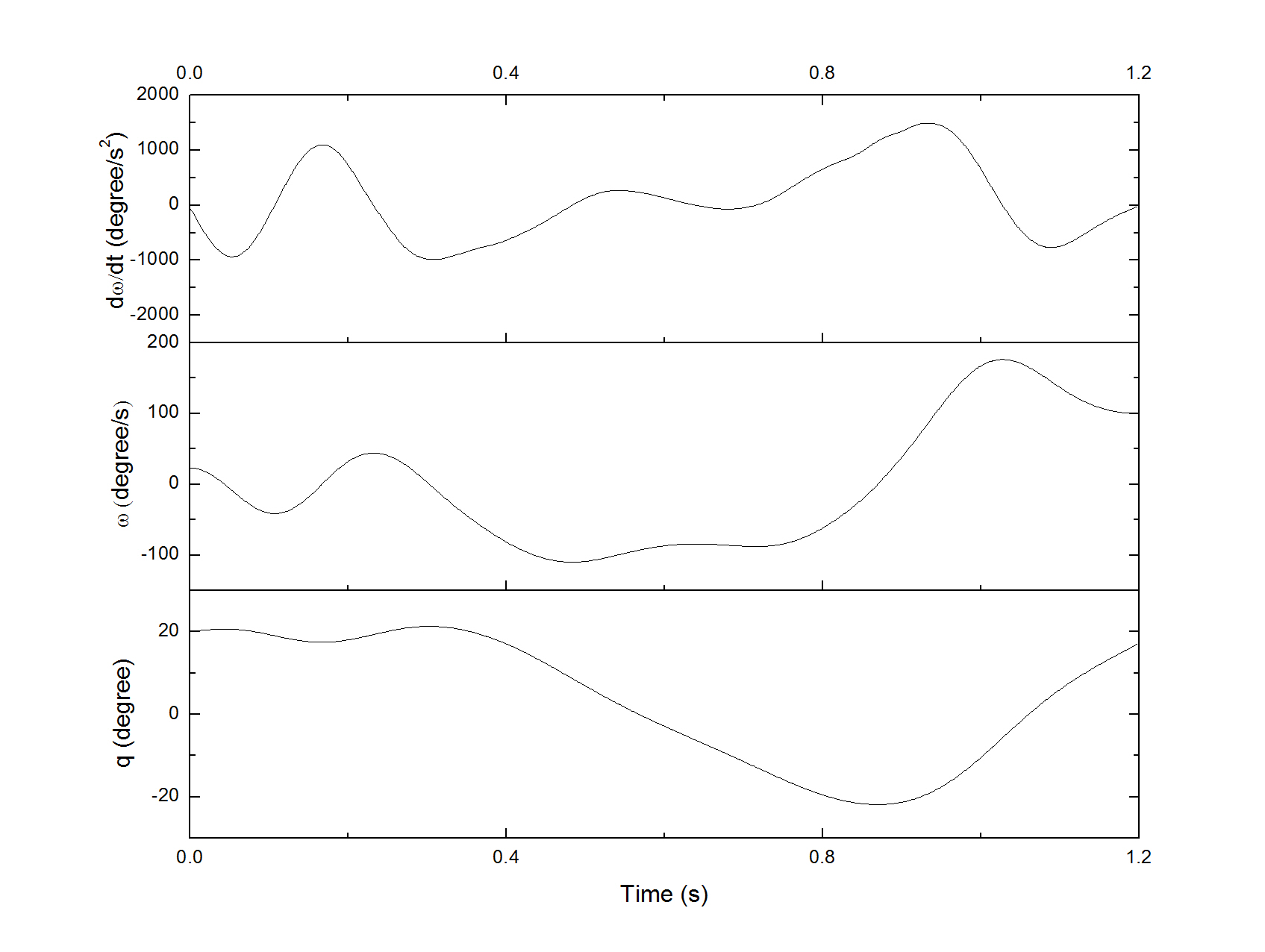}\\
  \caption{Coordinate (q), velocity ($\omega$), acceleration (d$\omega$/dt) of right hip flexion in the motion.}
  \label{Figure8}
 \end{figure}
Inverse dynamic analysis on the two models generates different results. Figure 9 shows the right hip flexion moments calculated from the two models. With the approximate model as yardstick, the error of the calculated right hip flexion moment of the scaled model has a mean of 1.89~Nm, which is 10.11\% of its mean absolute value. A total of 18 joint moments was calculated. The means of error percentage vary from 0.65\% to 12.68\%, with an average of 5.01\%. Relative data are shown in Table 1.
\begin{figure}[htbp]
  \centering
  \includegraphics[width=1\textwidth]{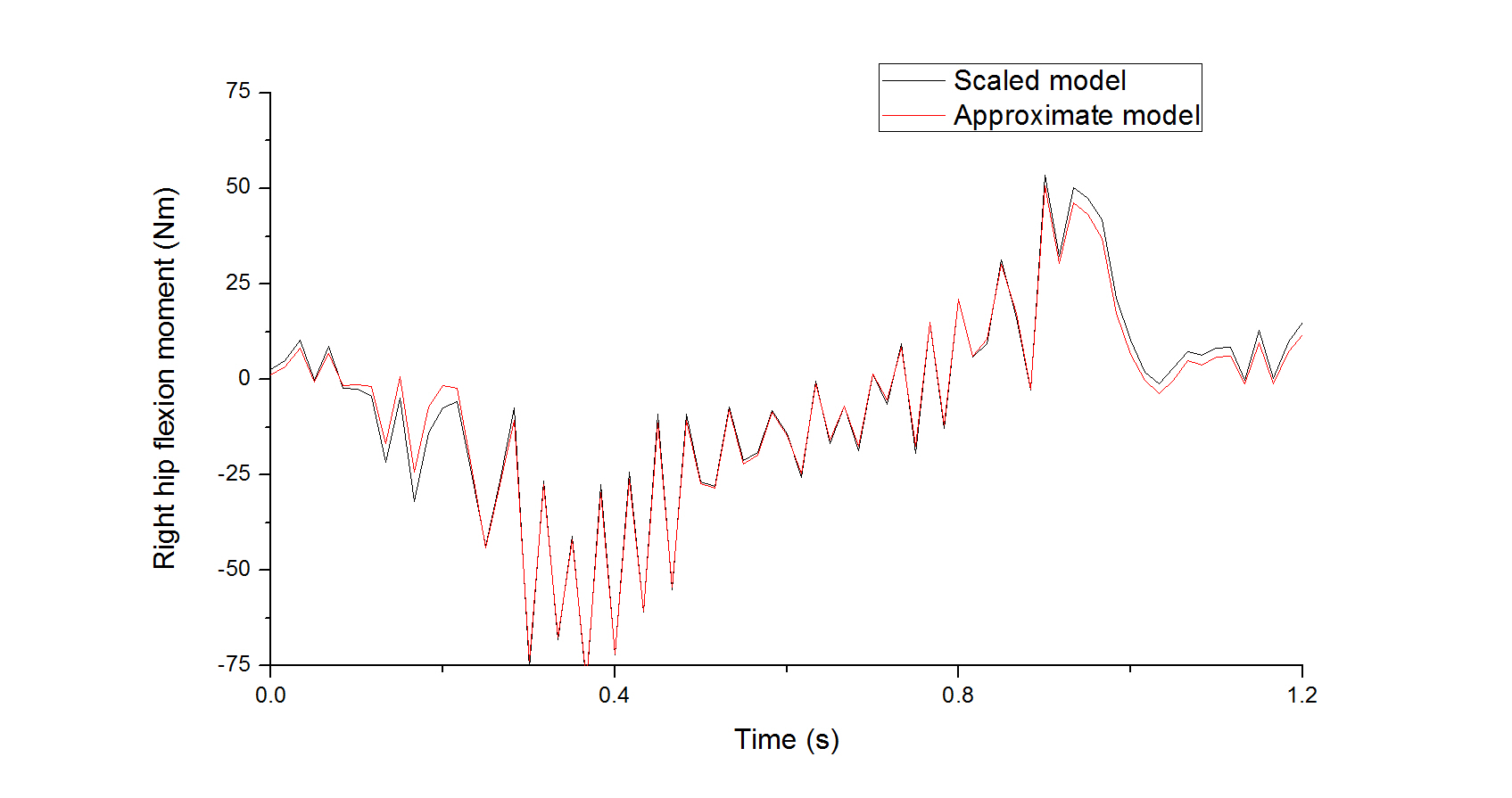}\\
  \caption{Right hip flexion moments calculated from the two models.}
  \label{moment}
 \end{figure} 

\begin{table}[htbp]
  \caption{Errors of the joint moments calculated from the scaled model, with respect to that from the approximate model}
  {\begin{tabular}{p{3.5cm}p{2.5cm}p{3cm}p{3cm}}
  \toprule
  {} & {\noindent{\textbf{Mean of error (Nm)}}} & {\noindent{\textbf{Mean of instant moment (Nm)}}} & {\noindent{\textbf{Mean of error percentage}}}\\
\midrule
Pelvis tilt &6.73&64.12&10.50\%\\
Pelvis list &4.06&37.78&10.75\%\\
Pelvis rotation & 1.02 & 17.78 & 5.74\% \\
Right hip flexion &1.89&18.73&10.11 \% \\
Right hip adduction &0.48&18.57&2.60 \% \\
Right hip rotation &0.07&3.47&2.13 \% \\
Right knee angle &0.65&19.21&3.40 \% \\
Right ankle angle &0.31&38.68&0.80 \% \\
Right subtalar angle &0.06&9.02&0.65 \%\\
Left hip flexion &2.27&17.90&12.68 \%\\
Left hip adduction &0.80&24.99&3.18 \%\\
Left hip rotation &0.10&4.28&2.23 \%\\
Left knee angle &0.83&19.37&4.28 \%\\
Left ankle angle &0.36&30.18&1.19 \%\\
Left subtalar angle &0.07&7.65&0.91 \%\\
Lumbar extension &5.47&60.85&9.00 \%\\
Lumbar bending &3.05&37.37&8.17 \%\\
Lumbar rotation &0.33&17.98&1.85 \%\\
\bottomrule
\end{tabular}}
 \label{tab3}
\end{table}
%%%%%%%%%%%%%%%%%%
\section{Discussions}
%%%%
\subsection{The use of 3D scan in the estimation of body segment masses}
%%%%
In previous researches, the inertial parameters of human body segment are usually determined by two means: (i) Applying predictive equations generated from database \cite{DeLeva1996},  (ii) Medical scanning of live subjects \cite{Lee2009}, and  (ii) Segments geometric modelling \cite{Davidson2008}. The use of the first one, as stated by Durkin \& Dowling (2003) \cite{Durkin2003}, is limited by its sample population. Furthermore, the difference in segmentation methods makes it difficult to combine various equations \cite{Pearsall1994a}. The second method, medical scanning, such as dual energy X-ray absorptiometry is more accurate in obtaining body segment inertial parameters \cite{Durkin2002}. But it is more expensive and time-consuming. 

In this study, body segment masses are estimated by segment density data and segment volume. 3D scan is used to estimate body segment volume. In this process, errors may merge from two aspects. 

First, it is assumed that density data of each segment is constant among humans. This assumption may bring errors. Traditional body composition method defines two distinct body compartments: fat and lean body (fat-free). Fat has a density of 0.90~g/ml, while lean body has a density of 1.10~g/ml \cite{Lukaski1987}. Subject’s body fat rate may influence the segment density.  However, the range of density variation is smaller than that of the mass distribution. Therefore, the use of density and volume may reduce the estimation error of segment mass. 

For example, in the current study, the thigh, with a volume of 8.35~l, holds a mass proportion that would vary from 9.74\% (all fat, density = 0.90~g/ml) to 11.90\% (fat-free, density = 1.10~g/ml). This range is much narrower than that found in previous research (from 9.2\% \cite{okada1996} to 14.47\% \cite{DeLeva1996}).  

Second, 3D scan is used to build up 3D geometric model and calculate segment volumes. Significant difference exists between volumes calculated and volumes measured by water displacement. To approximate the real volume, assumption is made that the 3D geometric model has the same volume distribution with that of the subject, which may bring error. 

In summary, as a simple and low-cost method of segment mass determination, the use of density data and 3D geometric model is more likely to reduce the estimation error. 3D scan is an easy way to construct a 3D geometric model, but attention should be payed to the model's volume errors. The method used in this study to approximate the real segment volumes with 3D scanned model needs to be examined in future researches.  
%%%%
\subsection{The  error and error significance of the proportionately scaled model }
%%%%
Proportional scaling is efficient to specific a generic model. In this study, relative errors of segment masses of the scaled model are between 4.06\% to 47.42\%. The error of torso mass is 4.09~kg, which takes up to 5.30\% of the overall body weight. In the following motion simulation, these errors bring about difference in the calculated joint moment. Means of the difference of calculated joint moments are from 3.65\% to 12.68\%. This suggests that a careful specification of segment masses will increase the accuracy of the dynamic simulation.
%%%%%%%%%%%%%%%%%%%%%%%%%%%%%%%%%%%%%
\section{Conclusions}
%%%%%%%%%%%%%%%%%%%%%%%%%%%%%%%%%%%%%
This study aims at estimating the errors and their influences on dynamic analysis caused by the scaling method used in OpenSim. A 3D scan is used to construct subject's 3D geometric model, according to which segment masses are determined.  The determined segment masses data is taken as the yardstick to assess OpenSim proportionately scaled model: errors are calculated, and  influence of the errors on dynamics analysis is examined.

As a result, the segment mass error reaches up to 5.31\% of the overall body weight (torso). Influence on the dynamic calculation has been found, with a average difference from 3.65\% to 12.68\% in the joint moments. 

Conclusions could be drawn that (i) the use of segment volume and density data  may be more accurate than mass distribution reference data in the estimation of body segment masses and (ii) a careful specification of segment masses will increase the accuracy of the dynamic simulation significantly. This current work is a study to determine inertial parameters of the human body segment in biomechanical simulation. It explores new, more precise and simpler ways to implement biomechanical analysis. This work is a step towards characterizing muscular capacities for the analysis of work tasks and predicting muscle fatigue.
%%%%%%%%%%%%%%%%%%%%%%%%%%%%%%%%%%%%%
\section{Acknowledge}
This work was supported by INTERWEAVE Project (Erasmus Mundus Partnership Asia-Europe) under Grants number IW14AC0456 and IW14AC0148, and by the National Natural Science Foundation of China under Grant numbers 71471095 and by Chinese State Scholarship Fund. The authors also thank D.Zhang Yang for his support.
%%%%%%%%%%%%%%%%%%%%%%%%%%%%%%%%%%%%%
%
% ---- Bibliography ----
\bibliographystyle{splncs.bst}
\bibliography{FBModelbib.bib}
\end{document}